\documentclass[reprint,showpacs,amssymb,amsmath,aps,pra]{revtex4-1}
\usepackage{graphics}
\usepackage{bm}
\usepackage{amsmath}
\usepackage{amssymb}
\begin{document}

\title{Measurements, quantum discord and parity in spin 1 systems}
\author{R. Rossignoli, J.M. Matera, N. Canosa}
\affiliation{Departamento de F\'{\i}sica-IFLP,
Universidad Nacional de La Plata, C.C. 67, La Plata (1900), Argentina}

\pacs{03.67.Mn, 03.65.Ud, 03.65.Ta}
\begin{abstract}
We consider the evaluation of the quantum discord and other related measures of
quantum correlations in a system formed by a spin $1$ and a complementary spin
system. A characterization of general projective measurements in such system in
terms of spin averages is thereby introduced, which allows to easily visualize
their deviation from standard spin measurements. It is shown that the
measurement optimizing these measures corresponds in general to a non-spin
measurement. The important case of states that commute with the total $S_z$
spin parity is discussed in detail, and the general stationary measurements for
such states (parity preserving measurements) are identified. Numerical and
analytical results for the quantum discord, the geometric discord and the one
way information deficit in the relevant case of a mixture of two aligned spin
$1$ states are also presented.
\end{abstract}
\maketitle

\section{Introduction}
There is presently a great interest in the investigation of quantum
correlations and ``quantumness'' in mixed states of composite quantum systems.
While in the case of pure states such correlations can be identified with
entanglement, the situation in mixed states is more complex, as separable
(i.e., non-entangled) mixed states, defined as convex mixtures of product
states \cite{RW.89}, can still exhibit signatures of quantum correlations, as
the different products may not commute. The interest has been further enhanced
by the existence of mixed state based quantum algorithms, such as that of Knill
and Laflamme (KL) \cite{KL.98}, able to achieve an exponential speed-up over
the classical algorithms with no entanglement \cite{DFC.05}. In contrast,
entanglement is essential for achieving exponential speed-up in pure state
based quantum computation \cite{JL.03}.

Consequently, alternative measures of quantum correlations for mixed states,
such as the quantum discord \cite{OZ.01,HV.01,Zu.03}, have recently received
much attention. Though coinciding with entanglement in pure states, discord
differs essentially from the latter in mixed states, being non-zero in most
separable states and vanishing just for ``classically correlated'' states,
i.e., states which are diagonal in a standard or conditional product basis. The
existence of a finite discord in the KL algorithm \cite{DSC.08} further
increased the interest on this measure. Other measures with similar properties
were also recently introduced
\cite{Lu.08,Mo.10,DVB.10,HH.05,RCC.10,SKB.11,Mo.11,GA.12}, including in
particular the geometric discord \cite{DVB.10}, which allows an easier
evaluation. Various fundamental properties
\cite{Mo.10,DVB.10,HH.05,RCC.10,SKB.11,Mo.11,FF.10,FC.11} and operational
interpretations
\cite{SKB.11,Mo.11,DG.09,MD.11,CC.11,PG.11,RRA.11,GC.12,DL.12,TG.12} of these
measures were recently unveiled. For instance, from the results of \cite{KW.04}
it follows that in a pure tripartite system $|\Psi_{ABC}\rangle$, the quantum
discord between $C$ and $A$ (as obtained due to a measurement in $C$) is the
entanglement of formation \cite{Be.96} between $A$ and $B$ plus the conditional
entropy $S(A|B)$ \cite{CC.11,MD.11,FC.11}. This entails that such discord
provides the entanglement consumption in the extended quantum state merging
scheme from $A$ to $B$  \cite{CC.11,MD.11}. Besides, states with non-zero
discord can be used, even if separable, to generate entanglement in the
protocols of \cite{SKB.11} or \cite{PG.11},  with the quantum discord and the
one-way information deficit \cite{HH.05,SKB.11} (a closely related quantity)
providing  the minimum partial and total distillable entanglement between the
measurement apparatus and the system after a von Neumann measurement on the
latter \cite{SKB.11}. Operational interpretations of the geometric discord were
also recently provided \cite{DL.12,TG.12}. See ref.\ \cite{Mo.11} for a recent
review.

A common feature of discord type measures is that they involve a difficult
minimization over a general local measurement on one of the system
constituents. Consequently, most evaluations were so far restricted to two
qubits (two spins $1/2$) or a qubit plus a complementary system, where the most
general projective measurement in the local qubit reduces to a standard spin
measurement and is hence easy to parameterize
\cite{OZ.01,DVB.10,DSC.08,AR.10,CRC.10,GA.11,RCC.11}. Closed evaluations in
gaussian systems with gaussian type measurements were also achieved
\cite{GP.10,AD.10}. Nonetheless, even for two qubits, general analytic
expressions are available just for the geometric discord \cite{DVB.10} and some
related measures \cite{RCC.11}. Here we will examine the
evaluation of the quantum discord (and related measures) between a spin $1$ and
a complementary spin system. This requires first a convenient characterization
of measurements in a spin $1$ system (a qutrit), since they are no longer
restricted to standard spin measurements as in the spin $s=1/2$ case, even when
considering just standard projective measurements. We provide in sec.\ II  a
simple description of such measurements in terms of spin averages, and show
that spin measurements are not optimum in general for spin $s\geq 1$, even if
the state is described in terms of basic spin observables.

We then analytically identify, in sec.\ III, the stationary projective
measurements for states exhibiting $S_z$ parity symmetry, an ubiquitous
symmetry present for instance in any non-degenerate eigenstate of spin arrays
with $XYZ$ couplings of arbitrary range in a transverse field \cite{RCM.09}
(for a pair of qubits such symmetry leads to the well-known $X$ states
\cite{AR.10}). This allows a considerable simplification of the problem of
discord evaluation in parity conserving systems. As application, we present
analytical results for the quantum discord, the geometric discord and the
one-way information deficit in the important case of a mixture of two aligned
spin 1 states. Such mixture represents the reduced state of {\it any} spin pair
in the ground state of $XYZ$ spin $1$ chains in the immediate vicinity of the
transverse factorizing field \cite{GAI.08,RCM.09}, so that present results
represent the universal limit of these quantities at such point. We also 
explicitly determine the projective measurements minimizing these quantities
for this state and show that they exhibit important differences. 
Conclusions are finally given in IV.

\section{Measurements in spin systems}
\subsection{General case}
We first consider a spin $s$ system, where we will denote with
$\bm{S}=(S_x,S_y,S_z)={\cal S}/\hbar$ the dimensionless angular momentum and
$|m\rangle$ the eigenstates of $S_z$ (standard basis). Spin measurements are
measurements in a basis of eigenstates
$|m_{\bm{k}}\rangle=e^{-i\bm{\theta}_{\bm{k}}\cdot\bm{S}}|m\rangle$ of the spin
component $\bm{k}\cdot\bm{S}$ along the direction of a unit vector $\bm{k}$,
and are then specified by just two real parameters which determine its
orientation. For $s\geq 1$ these measurements are, however, only a  particular
case of complete projective measurement (von Neumann measurement), i.e., those
defined by a complete set of rank $1$ orthogonal projectors. The latter are
determined by a general unitary transformation of the $S_z$ eigenstates,
\begin{equation}
|m_U\rangle=U|m\rangle\,,\label{1}
\end{equation}
and depend therefore on $d(d-1)$ real parameters, with $d=2s+1$ ($U=e^{iH}$,
with $H$ hermitian, depends on $d^2$ real parameters, but just $d^2-d$ are
sufficient to determine the set of projectors $\{\Pi_m^U=|m_U\rangle\langle
m_U|\}$ defining the measurement, as the phase of each $|m_U\rangle$ is
irrelevant). The  states (\ref{1}) are the eigenstates of the operator
$S^U_z=US_zU^\dagger$, which in general is no longer a linear combination of
the original $S_\mu$ ($\mu=x,y,z$). Such measurements can, nonetheless, be
regarded as measurements of a generalized spin $S^U_z$ (the algebra
$[S^U_\mu,S^U_\nu]=i\epsilon_{\mu\nu\sigma}S^U_{\sigma}$ still holds), and can
be implemented as measurements in the standard $S_z$ basis preceded by a single
{\it qudit} gate $U^\dagger$.

A first glimpse into the nature of these measurements can be attained through
the set of vectors
\begin{equation}
\langle \bm{S}\rangle_{m_U}=\langle m_U|\bm{S}|m_U\rangle\,,\label{2}
\end{equation}
which, in contrast with the case of a spin measurement ($\langle
\bm{S}\rangle_{m_{\bm{k}}}=m\bm{k}$), i) may have any length between 0 and $s$
and ii) are not necessarily collinear. Nonetheless, since $\bm{S}$ is
traceless, they always sum to zero:
\begin{equation} \sum_{m}\langle
\bm{S}\rangle_{m_U}=\bm{0}\,.\label{S0}\end{equation}
While not fully identifying the measurement, the
set of averages (\ref{2}) allow a rapid visualization of its  deviation from a
standard spin measurement: if $\langle \bm{S}\rangle_{m_U}=m\bm{k}$ for
$m=-s,\ldots,s$, it is clearly a spin measurement along $\bm{k}$ due to the
orthogonality of the basis.

\subsection{Spin 1 systems}
\begin{figure}
\vspace*{-0.25cm}

\centerline{\scalebox{.65}{\includegraphics{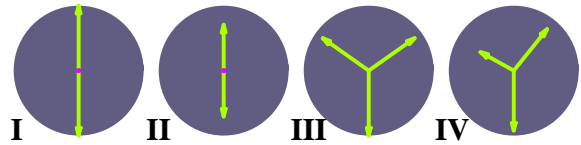}}}
 \vspace*{-0.25cm}

\caption{(Color online) Representation of measurements in a spin 1 system
through the spin expectation values in the basis states. I: Spin measurement
along $z$. II: Collinear measurement, determined by the definite parity states
(\ref{st1}) or (\ref{st1p}) ($\beta=0$ in (\ref{str1})). III: $Y$-type
measurement, determined by the basis (\ref{str3})--(\ref{str4}) ($\beta=\pi/4$
in (\ref{str1})). IV: General measurement, determined by basis
(\ref{str1})--(\ref{str3}).}
 \label{f1}\vspace*{-0.35cm}
 \end{figure}

In the case of a spin $1$ system ($d=3$), Eq.\ (\ref{S0}) entails that the
three vectors (\ref{2}) are {\it coplanar}. Moreover, the operators $S_z^U$ are
at most  quadratic functions of the $S_\mu$, as any operator in such system can
be written as a linear combination of the three $S_\mu$ and the six operators
$(S_\mu S_\nu+S_\nu S_\mu)/2$. For example, a non-spin measurement in such
system is provided by the states
$|m_{\alpha}\rangle=e^{-i\alpha(S_xS_y+S_yS_x)}|m\rangle$, i.e.,
\begin{equation}
|\pm 1_{\alpha}\rangle=\cos\alpha\,|\pm 1\rangle\pm \sin\alpha\,|\mp
1\rangle,\;|0_\alpha\rangle=|0\rangle\,,\label{st1}
 \end{equation}
which satisfy $S_z^\alpha|m_\alpha\rangle=m|m_\alpha\rangle$ with
\[S_z^\alpha=S_z\cos 2\alpha+(S_x^2-S_y^2)\sin 2\alpha\,.\]
They lead to
\[\langle \bm{S}\rangle_{\pm 1_\alpha}=(0,0,\pm\cos\,2\alpha),\;\;
 \langle \bm{S}\rangle_{0_\alpha}=\bm{0}\,,\]
and hence to the second plot in Fig.\ \ref{f1}: The vectors  $\langle
\bm{S}\rangle_{m_\alpha}$ are still collinear but $|\langle \bm{S}\rangle_{\pm
1_\alpha}|\leq 1$. Moreover, for $\alpha=\pi/4$,
$\langle\bm{S}\rangle_{m_\alpha}=\bm{0}$ $\forall$ $m$, showing that the
average spin may vanish in {\it all} elements of the basis: In this case $|\pm
1_\alpha\rangle=(|\pm 1\rangle\pm |\mp 1\rangle)/\sqrt{2}$ become the zero
eigenstates of $S_y$ and $S_x$ respectively, which form together with
$|0\rangle$ an orthonormal basis.

The most general basis (disregarding global phases and permutations) leading to
collinear averages along $z$ for $s=1$ can be obtained by rotating the states
(\ref{st1}) around the $z$ axis, which leads to states
\begin{equation}
|m^\phi_\alpha\rangle=e^{-i\phi S_z}|m_\alpha\rangle\,.\label{st1p}
 \end{equation}
These are the most general states with {\it definite $S_z$ parity}:
\begin{equation}
P_z|m_\alpha^\phi\rangle=(-1)^{m+1}|m_\alpha^\phi\rangle\,,
\;\;P_z\equiv  e^{i\pi(S_z+1)}\,.
\end{equation}

We now show that the six parameters specifying a general projective measurement
in a spin $1$ system can be decomposed into three angles
$(\alpha,\beta,\gamma)$ which determine the ``intrinsic'' plot of vectors
$\langle\bm{S}\rangle_{m_U}$ (and hence the type of measurement), plus three
angles $(\psi,\theta,\phi)$ which determine the orientation of this plot and of
the ensuing states. Assuming first $\langle\bm{S}\rangle_{m_U}\neq\bm{0}$ for
some $m$, we choose the ``intrinsic'' $z$ axis in the direction of this vector.
A state $a|1\rangle+b|0\rangle+c|-1\rangle$ giving rise to  $\langle
\bm{S}\rangle=(0,0,\langle S_z\rangle)$ should satisfy $b\bar{a}+\bar{b}c=0$,
which implies $b=0$ if $\langle S_z\rangle\neq 0$ ($|a|\neq |c|$). Discarding
total phases, the most general orthonormal basis containing such state is then
\begin{eqnarray}
&&\!\!\!\!\!|{^{1_{\bm{r}}}_{0_{\bm{r}}}}\rangle={^{\cos\beta}_{\sin\beta}}
(e^{-i\phi_0}\cos\alpha|1\rangle+e^{i\phi_0}\sin\alpha|-1\rangle)
\mp {^{\sin\beta}_{\cos\beta}}e^{-i\gamma}|0\rangle\,,  \label{str1}\\
&&\!\!\!\!\!|-1_{\bm{r}}\rangle=-e^{-i\phi_0}\sin\alpha|1\rangle
+e^{i\phi_0}\cos\alpha|-1\rangle\,.
\label{str3}
\end{eqnarray}
where $\bm{r}\equiv (\alpha,\beta,\gamma)$.
These states lead in general to non-collinear spin averages of different
lengths (plot IV in Fig.\ \ref{f1}). Choosing $\phi_0$ such that the diagram
lies in the intrinsic $x,z$ plane ($\langle S_y\rangle_{m_{\bm{r}}}=0$
$\forall$ $m$), we obtain $\tan\phi_0=\tan\gamma\tan(\pi/4-\alpha)$ and
\begin{eqnarray}
\langle \bm{S}\rangle_{^{1_{\bm{r}}}_{0_{\bm{r}}}}
&=&(\mp\sin 2\beta{\textstyle\sqrt{(1+\cos 2\gamma\sin 2\alpha)/2}},
0,{^{\cos^2\beta}_{\sin^2\beta}}\cos 2\alpha)\,,\nonumber\\
\langle \bm{S}\rangle_{-1_{\bm {r}}}&=&(0,0,-\cos
 2\alpha)\,.\label{vms}\end{eqnarray}
Hence, $(\alpha,\beta,\gamma)$  determine respectively
$\langle\bm{S}\rangle_{-1_{\bm{r}}}$ and the components of $\langle
\bm{S}\rangle_{{0}_{\bm{r}}}$ parallel and orthogonal to
$\langle\bm{S}\rangle_{-1_{\bm{r}}}$. Eqs.\ (\ref{vms}) also show that the
angle between vectors $\langle \bm{S}\rangle_{m_{\bm{r}}}$ always {\it exceeds}
$\pi/2$: $\!\langle\bm{S}\rangle_{m_{\bm{r}}}\cdot\langle
\bm{S}\rangle_{m'_{\bm{r}}}\leq 0$ if $m\neq m'$, vanishing just if one average
is zero. The states (\ref{str1})--(\ref{str3}) can be written as
$|m_{\bm{r}}\rangle=e^{i(\gamma(S_z^2-1)
-\phi_0S_z)}e^{-i\alpha(S_xS_y+S_yS_x)} e^{i\frac{\beta}{\sqrt{2}}
(S_y+S_yS_z+S_zS_y)}|m\rangle$.

The most general orthonormal basis is then obtained by applying a general
rotation $e^{-i\psi S_z}e^{-i\theta S_y}e^{-i\phi S_z}$ to this basis. This
also includes the case $\langle \bm{S}\rangle_{m_U}=\bm{0}$ $\forall$ $m$,
since such basis are always formed by the zero eigenstates of the components of
$\bm{S}$ along three orthogonal directions: For a state
$a|1\rangle+b|0\rangle+c|-1\rangle$, the condition $\langle
\bm{S}\rangle=\bm{0}$ implies $b\bar{a}+\bar{b}c=0$ and $|a|=|c|$. It is then
the eigenstate with zero eigenvalue of $\bm{k}\cdot\bm{S}$, with (assuming,
with no loss of generality, $b$ real and $c=-\bar{a}$) $\bm{k}=(-\sqrt{2} {\rm
Re}(a),\sqrt{2}{\rm Im}(a),b)$. Orthogonality of the basis states then implies
that of the associated vectors $\bm{k}$ (as
$\bm{k}\bm{k}'=a\bar{a}'+bb'+\bar{a}a'$). Hence, these basis can be obtained,
for instance, through a suitable rotation of the intrinsic case
$\alpha=\pi/4,\beta=\gamma=0$, where (\ref{str1})--(\ref{str3}) reduce to the
zero eigenstates of $S_y$, $S_z$ and $S_x$.

We may then set $\alpha,\beta\in[0,\pi/4]$ and $\gamma\in(-\pi/2,\pi/2]$ in
(\ref{str1})--(\ref{str3}), as other values  can be mapped to these ranges
after suitable rotations (disregarding total phases). Notice that if
$\gamma\in(0,\pi/2)$ and $\beta\neq 0$, the values $\pm \gamma$ lead to
inequivalent and conjugate basis (as $\phi_0(-\gamma)=-\phi_0(\gamma)$), but
the same set of spin averages. The definite parity states (\ref{st1}) are
recovered for $\beta=\gamma=0$.

Another relevant case is $\beta=\pi/4$ in (\ref{str1}), where
 \begin{equation}|{^{1_{\bm{r}}}_{0_{\bm{r}}}}\rangle=
{\textstyle\frac{1}{\sqrt{2}}}(e^{-i\phi_0}\cos\alpha|1\rangle+e^{i\phi_0}
\sin\alpha|-1\rangle\mp e^{i\gamma}|0\rangle)\,,\label{str4}
 \end{equation}
satisfy $P_z|^{1_{\bm{r}}}_{0_{\bm{r}}}\rangle=
|^{0_{\bm{r}}}_{1_{\bm{r}}}\rangle$, with
$P_z|-1_{\bm{r}}\rangle=|-1_{\bm{r}}\rangle$, such that parity also leaves this
basis (i.e., the set of states) {\it invariant}. This also implies  $\langle
1_{\bm{r}}|S_z|1_{\bm{r}}\rangle= \langle 0_{\bm{r}}|S_z|0_{\bm{r}}\rangle$,
entailing a {\it symmetric} $Y$-type spin diagram (plot III in Fig.\ \ref{f1}).
For $\alpha=\pi/4$, the $Y$ reduces to an horizontal line and the states
(\ref{str3})--(\ref{str4}) become, for $\gamma=\phi_0=0$, eigenstates of $S_x$
(as $\langle \bm{S}\rangle_{^{0_{\bm{r}}}_{1_{\bm{r}}}}=(\pm 1,0,0)$). The {\it
fully symmetric case} $\beta=\pi/4$, $\gamma=0$, $\sin 2\alpha=1/3$, where
$|\langle \bm{S}\rangle_{m_{\bm{r}}}|^2=8/9$ $\forall$ $m$, leads to the {\it
maximum} total squared spin length:  $L_S^2=\sum_m |\langle
\bm{S}\rangle_{m_U}|^2=8/3$, larger than the value $L_S^2=2$ obtained for a
spin measurement.

\section{Evaluation of Quantum discord and related measures}
\subsection{General Case}
Let us now consider the evaluation of the quantum discord between $n-1$
arbitrary spins $S^i$ (system $A$) and a spin $s$ (system $B$), as obtained due
to a local complete projective measurement $M_B=\{\Pi_m^U\}$ on system $B$. If
initially in a state $\rho_{AB}$, the state of the total system after an unread
measurement $M_B$ becomes
\begin{equation}
\rho'_{AB}=\sum_m (I_A\otimes \Pi_m^U)\rho_{AB}(I_A\otimes\Pi_m^U)\,.\label{rhop}
\end{equation}
For a local measurement of this type, the quantum discord \cite{OZ.01,HV.01}
can be expressed in terms of (\ref{rhop}) as
\begin{equation}
D^{B}(\rho_{AB})=\mathop{\rm Min}_{M_B}
[S(\rho'_{AB})-S(\rho'_B)]-[S(\rho_{AB})-S(\rho_{B})]\,,\label{D}
\end{equation}
where $S(\rho)=-{\rm Tr}\,\rho\log\rho$ is the von Neumann entropy and
$\rho_B={\rm Tr}_A\,\rho_{AB}$ the reduced state of $B$. It can then be
considered as the minimum increase of the conditional entropy
$S(A|B)=S(A,B)-S(B)$ due to such measurements, and  is a non-negative quantity
\cite{OZ.01,HV.01}. For a pure state ($\rho_{AB}^2=\rho_{AB}$) it becomes the
entanglement entropy $S(\rho_B)=S(\rho_A)$, as in this case $S(\rho_{AB})=0$
and $S(\rho'_{AB})=S(\rho'_A)$ $\forall$ $M_B$ of this form. However, for a
mixed state $D^B(\rho_{AB})$ vanishes just for classically correlated states
with respect to $B$, i.e., states of the  form (\ref{rhop}) (a particular case
of separable state), which are diagonal in a conditional product basis
$\{|\nu_m\rangle\otimes|m_U\rangle\}$ and remain hence unchanged under a
particular von Neumann measurement in $B$. Eq.\ (\ref{D}) actually provides an
upper bound to the quantum discord obtained with general POVM measurements,
although results for two-qubits indicate that the difference is very small
\cite{Mo.11}.

We will also consider here the minimum generalized information loss due an
unread local measurement of the previous type \cite{RCC.10,RCC.11},
 \begin{equation}
 I^B_f(\rho_{AB})=\mathop{\rm Min}_{M_B} S_f(\rho'_{AB})-S_f(\rho_{AB})\,,
\label{If}
\end{equation}
where $S_f(\rho)={\rm Tr}\,f(\rho)$ denotes a general entropic form, with $f$ a
smooth strictly concave function satisfying $f(0)=f(1)=0$ \cite{CR.02}. Like
$D^B$, it can be shown \cite{RCC.10} that $I_f^B(\rho_{AB})\geq 0$ for any such
$f$ and $\rho_{AB}$, with $I_f^B(\rho_{AB})$ becoming the generalized
entanglement entropy $S_f(\rho_B)=S_f(\rho_A)$ for a pure state, while for a
general mixed state it vanishes just for states of the general form
(\ref{rhop}), i.e. states diagonal in a conditional product basis. Other
properties, including  the evaluation of $I_f^B$ for any $f$ in some specific
states (mixture of a pure state with a maximally mixed state, Bell-diagonal
states, etc.), were discussed in \cite{RCC.10,RCC.11}.

Eq.\ (\ref{If}) contains as particular cases two important measures: If
$f(\rho)=-\rho\log\rho$, $S_f(\rho)$ is the von Neumann entropy and  Eq.\
(\ref{If}) becomes \cite{RCC.10} the {\it one way information deficit} from $B$
to $A$ \cite{HH.05,SKB.11}. This quantity is closely related to the quantum
discord (\ref{D}), coinciding with it when the minimizing measurement is the
same for both quantities and such that $\rho'_{B}=\rho_{B}$ (this occurs for
instance when $\rho_B$ is maximally mixed, as in Bell diagonal states). It also
reduces to the  standard entanglement entropy $S(\rho_A)=S(\rho_B)$ for pure
states.  The one-way information deficit has been interpreted as the amount of
information that cannot be localized through a classical communication channel
from $B$ to $A$ \cite{HH.05,SKB.11}, and as previously stated, an operational
interpretation as the minimum distillable entanglement between the system and
the measurement apparatus, was recently provided \cite{SKB.11}.

On the other hand, if $f(\rho)=f_2(\rho)\equiv\rho(1-\rho)$, $S_f(\rho)$
becomes the so called {\it linear entropy} $S_2(\rho)=1-{\rm Tr}\,\rho^2$ and
Eq.\ (\ref{If}) becomes
\[I_2^B(\rho_{AB})=\mathop{\rm Min}_{M_B}{\rm Tr}\,(\rho_{AB}^2-{\rho'}_{\!AB}^2)\,.\]
This quantity is identical \cite{RCC.10} with the {\it geometric measure of
discord} \cite{DVB.10}, the latter defined as the minimum squared
Hilbert-Schmidt distance from $\rho_{AB}$ to a classically correlated state:
$I_2^B={\rm Min}_{\rho'_{AB}}||\rho_{AB}-\rho'_{AB}||^2$, where $||O||^2={\rm
Tr}\,O^\dagger O$ and $\rho'_{AB}$ is a state diagonal in a conditional product
basis with respect to $B$. In comparison with the previous measures, the
geometric discord offers the advantage of an easier evaluation (yet vanishing
for the same type of states), as the calculation of ${\rm Tr}\,\rho^2$ does not
require the explicit knowledge of the eigenvalues of $\rho$. An analytic
expression for general two qubit states was in fact provided in \cite{DVB.10},
while its extension to $2\otimes d$ systems was given in \cite{GA.12}. An
operational interpretation related with the fidelity and performance of remote
state preparation \cite{BB.01} (a variant of the teleportation protocol) has
also been recently provided \cite{DL.12,TG.12}. Besides, the geometric discord
for  a $2\otimes d$ system can be measured or estimated with direct
non-tomographic methods \cite{GA.12,JZ.12,PM.12}, which provide an
experimentally accessible scheme. For pure states $\rho_{AB}$, the geometric
discord becomes proportional to the square of the concurrence \cite{Ca.03}
$C_{AB}=\sqrt{2(1-{\rm Tr}\rho_B^2)}$.

The general stationary condition for Eq.\ (\ref{If}) (a necessary condition for
the minimizing measurement) reads \cite{RCC.11}
\begin{equation}
 \Delta_f^B\equiv{\rm Tr}_{A}[f'(\rho'_{AB}),\rho_{AB}]=0\,,
\label{stat}
\end{equation}
where $f'$ denotes the derivative of $f$. In the case of the quantum discord
(\ref{D}), an additional term $-[f'(\rho'_B),\rho_B]$
should be added to (\ref{stat}) to account for the
local terms in (\ref{D}), leading to the modified equation
\cite{RCC.11}
\begin{equation}
\Delta_D^B\equiv{\rm Tr}_{A}[f'(\rho'_{AB}),\rho_{AB}]
-[f'(\rho'_B),\rho_B]=0\,,\label{statd}
\end{equation}
where $f(\rho)=-\rho\log\rho$. Since $\Delta_f^B$ and $\Delta_D^B$ are
antihermitian local operators with zero diagonal elements in the measured basis
\cite{RCC.11}, they lead  to $d(d-1)/2$ complex equations, which determine
suitable values of the $d(d-1)$ real parameters defining the measurement in a
$d$ dimensional system $B$. They can be solved, for instance, with the gradient
method. It is then clear that standard spin measurements, defined by just two
real parameters, will not satisfy in general Eq.\ (\ref{stat}) or (\ref{statd})
for $s>1/2$, and hence cannot be minimum in general. In the spin $1$ case,
Eqs.\ (\ref{stat}) and (\ref{statd}) lead to 6 real equations which determine
suitable values of $(\alpha,\beta,\gamma)$ and the three rotation angles.

\subsection{States with $S_z$ parity symmetry and parity preserving measurements}
Let us now examine the important case where $\rho_{AB}$ commutes with the total
$S_z$ parity,
\begin{equation}
[\rho_{AB},P_z^{AB}]=0\,,\;\;P_z^{AB}=P_z^A\otimes
P_z^B\,,\label{rhpz}
 \end{equation}
where $P_z^A=\otimes_{i=1}^{n-1} e^{i\pi (S_z^i-S^i)}$. This is an ubiquitous
symmetry. For instance, general $XYZ$ type couplings of arbitrary range between
spins in a transverse field, not necessarily uniform, lead to a Hamiltonian
\begin{equation}
H=\sum_i b_i S^i_z-\sum_{i,j}\sum_{\mu=x,y,z}J^\mu_{ij}S^i_\mu S^j_\mu\,,
 \label{H}\end{equation}
which clearly satisfies $[H,P^{AB}_z]=0$, irrespective of the geometry and
dimension of the array. The same holds even if terms $\propto S_x^i S_y^j$ are
also present. Hence, any non-degenerate eigenstate of $H$, as well as the
thermal state $\rho_{AB}\propto\exp[-\beta H]$, will fulfill Eq.\ (\ref{rhpz}).
Moreover, if Eq.\ (\ref{rhpz}) holds, parity is also preserved at the {\it local}
level, i.e. $[\rho_B,P_z^B]=0$, as the partial trace involves just diagonal
elements in the complementary system $A$. The reduced state of any subgroup of
spins will then also commute with the corresponding local $S_z$ parity.

We also add that any system described by a  Hamiltonian containing just
quadratic terms $\propto$ $P_i P_j$, $Q_i Q_j$ and $Q_i P_j$ in standard
coordinates and momenta $Q_i=\frac{b_i+b^\dagger_i}{\sqrt{2}}$,
$P_i=\frac{b_i-b_i^\dagger}{\sqrt{2} i}$, with $b_i, b^\dagger_i$  boson
operators ($[b_i,b_j^\dagger]=\delta_{ij}$, $[b_i,b_j]=0$), does commute with
the boson number parity $P_N=e^{i\pi N}$, where $N=\sum_i b^\dagger_i b_i$.
Hence, when restricted to a finite $N$ subspace, (i.e., $b^\dagger_i b_i\leq
N_{\rm max}$), such system is equivalent to a spin like system whose
Hamiltonian commutes with the corresponding $S_z$ parity, defining
$S^i_{z}=b^\dagger_i b_i-N_{\rm max}/2$.

For an arbitrary $\rho_{AB}$ satisfying Eq.\ (\ref{rhpz}), parity  will be
preserved by the measurement $M_B$, i.e.,
\begin{equation}
 [\rho'_{AB},P_z^{AB}]=0\,,\label{rhppz}
\end{equation}
when $P_z^B \Pi_m^U P_z^B=\Pi_m^U$  $\forall$ $m$ {\it and also when} $P_z^B
\Pi_m^U P_z^B=\Pi_{m'}^U$, where $\Pi_{m'}^U$ is another element of the set of
local projectors, as in both cases the set will remain invariant: $\{P_z^B
\Pi_m^U P_z^B\}=\{\Pi_m^U\}$. The last case corresponds to
$P_z^B|m_U\rangle\propto |m'_U\rangle$, and since $(P_z^B)^2=I_B$, such basis
can contain just pairs permuted by $P_z^B$ and isolated eigenstates of $P_z^B$.
For a spin $1$ system, parity will then be preserved for type II as well as
type III measurements, i.e., those based on the states
(\ref{st1})--(\ref{st1p})
 or (\ref{str3})--(\ref{str4}).

If Eqs.\ (\ref{rhpz})--(\ref{rhppz}) hold, the commutator
in (\ref{stat}) will also commute with $P_z^{AB}$, implying
\begin{equation}
 [\Delta_f^B,P_z^B]=0\,,\;\;\;[\Delta_D^B,P_z^B]=0\,.\label{DP}
\end{equation}
This ensures the existence of parity preserving measurements satisfying Eq.\
(\ref{stat}) or (\ref{statd}), as the number of independent elements which have
to vanish is reduced by (\ref{DP}), matching exactly the reduced number of free
parameters defining such measurements (essentially $\approx d(d-1)/2$). For
instance, in the spin $1$ case and for type II measurements, Eq.\ (\ref{DP})
implies $(\Delta_f^B)_{m,0}=0$ in the measurement basis and Eq.\ (\ref{stat})
reduces to a single complex equation ($(\Delta_f^B)_{-1,1}=0$) determining
$\alpha,\phi$. For type III measurements, Eq.\ (\ref{DP}) implies
$(\Delta_f^B)_{0,1}$ imaginary and $(\Delta_f^B)_{-1,0}=(\Delta_f^B)_{-1,1}$ in
the measured basis, and Eq.\ (\ref{stat}) leads to one real and one complex
equation, which determine $\alpha,\gamma,\phi$. As there is a maximum and a
minimum of $I_f^B$ within these measurements, solutions are ensured. Moreover,
if $\Delta_f^B$ is {\it real} in the standard basis, as  occurs for instance
when $\rho_{AB}$ and all $\Pi_m^U$ are real in such basis ($\phi=\gamma=0$),
Eq.\ (\ref{DP}) {\it reduces to a single real equation in both measurements:}
\begin{equation}(\Delta_f^B)_{-1,1}=0\,,\label{D11}\end{equation}
which determines the optimum  $\alpha$. These arguments also apply for
$\Delta_D^B$, leading to $(\Delta_D^B)_{-1,1}=0$ in the real case.

Parity preserving measurements are then strong candidates for providing the
actual minimum of $D^B$ or $I_f^B$, although ``parity breaking'' solutions of
(\ref{stat}) may also exist. The latter are {\it degenerate}, as the sets
$\{\Pi_m^U\}$ and $\{P_z^B \Pi_m^U P_z^B\}$ will lead to the {\it same} values
of $D^B$ and $I_f^B$ when (\ref{rhpz}) holds. Note also that parity preserving
spin measurements are just those along $z$ or an axis perpendicular to $z$
(where $P_z|m_{\bm{k}}\rangle\propto|-m_{\bm{k}}\rangle$) and do not have
enough parameters for satisfying Eq.\ (\ref{stat}) if $s\geq 1$. In the real
case, just those along $x$, $y$ or $z$ will lead in general to a real
$\rho'_{AB}$ and no continuous free parameter is left.

\subsection{Application}
As illustration, we consider a bipartite state formed by the mixture of two
aligned spin $1$ states,
\begin{equation}
\rho_{AB}={\textstyle\frac{1}{2}}(|\theta\theta\rangle\langle\theta\theta|
+|-\theta-\theta\rangle\langle-\theta-\theta|)\,,
 \label{rhot}
 \end{equation}
where $|\theta\rangle\equiv e^{-i\theta S_y}|1\rangle=|1_{\bm{k}}\rangle$ is
the state with maximum spin along $\bm{k}=(\sin\theta,0,\cos\theta)$ (a coherent
state). As $P_z|\theta\rangle=|-\theta\rangle$, Eq.\ (\ref{rhot})  fulfills
Eq.\ (\ref{rhpz}). This state arises, for instance, as the reduced state of
{\it any} spin pair in the fixed parity states
\begin{equation}|\Psi_{\pm}\rangle=\frac{
|\theta\ldots\theta\rangle\pm|-\theta\ldots
-\theta\rangle}{\sqrt{2(1\pm\langle -\theta|\theta\rangle^n)}}
 \label{psi}\,,\end{equation}
if small overlap terms $\propto \langle -\theta|\theta\rangle^{n-2}$  are
neglected ($\langle -\theta|\theta\rangle=\cos^{2s}\theta$) \cite{CRC.10}. Such
states are the {\it exact} ground states of an $XYZ$ spin chain described by
(\ref{H}) in the immediate vicinity of the transverse factorizing field
\cite{RCM.09}, existing in the case of fixed anisotropy
$\chi=\frac{J^y_{ij}-J^z_{ij}} {J^x_{ij}-J^z_{ij}}=\cos^2\theta$ $\forall$ $i,j$ for
$|J^y_{ij}|\leq J^x_{ij}$, {\it irrespective} of the geometry or coupling
range.

The state (\ref{rhot}) is {\it separable} (a convex mixture of product states
\cite{RW.89}) $\forall$ $\theta$, but classically correlated just for
$\theta=0$ or $\pi/2$ (where $\langle -\theta|\theta\rangle=0$). Accordingly,
$D^B(\rho_{AB})$ and $I_f^B(\rho_{AB})$ will be non zero just for
$\theta\in(0,\pi/2)$. As the state is symmetric, we have $D^B=D^A\equiv D$ and
$I_f^B=I_f^A\equiv I_f$. As before, we will consider just von Neumann type
local projective measurements $M_B$.

\begin{figure}[t]
\vspace*{-0.5cm}

\centerline{\scalebox{.5}{\includegraphics{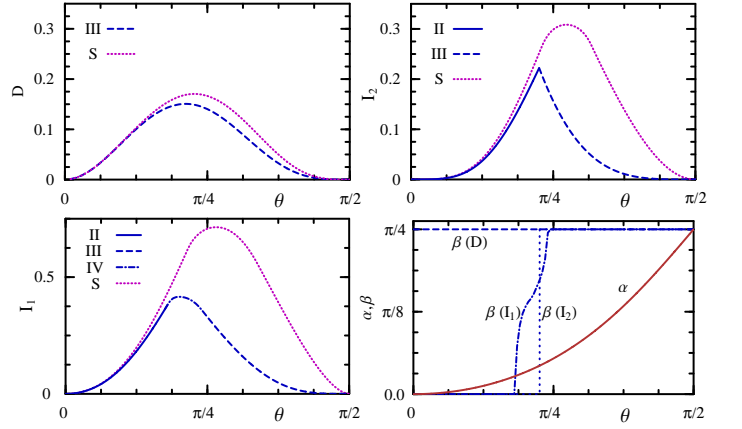}}}
 \vspace*{-0.5cm}

\caption{(Color online) The quantum discord $D$ (top left), the geometric
discord $I_2$ (top right) and the one way information deficit $I_1$ (bottom
left) of the mixture of aligned states (\ref{rhot}) as a function of $\theta$
for spin $s=1$. The dotted lines depict the result obtained with a spin
measurement, the other curves the actual minimum, obtained with the indicated
measurement (see Fig.\ \ref{f1}). The bottom right panel depicts the angles
characterizing the minimizing measurement. $D$ is minimized by a parity
preserving type III measurement $\forall$ $\theta$ whereas $I_2$ changes from
type II to III at $\theta=\theta_c$, and $I_1$ changes from II to III through a
small crossover region where a parity breaking type IV measurement is
preferred. The angle $\alpha$ is the same for all quantities (Eq.\ (\ref{al})).
Normalization is such that $D=I_2=I_1=1$ for a Bell state.}
 \label{f2} \end{figure}

Results for the quantum discord $D$, the geometric discord $I_2$ and the
one-way information deficit (denoted here as $I_1$) are shown in Fig.\
\ref{f2}. It is first confirmed that minimization over spin measurements
provides just an upper bound to the actual value of these quantities, being
nonetheless a good approximation for small $\theta$. The qualitative behavior
of these three quantities is similar (they are all maximum for $\theta$
slightly below $\pi/4$), but important differences in the minimizing
measurement do arise. While $D$ is minimized by a real ($\phi=\gamma=0$) type
III measurement $\forall$ $\theta\in(0,\pi/2)$, leading to a smooth curve,
$I_2$ prefers a real type II (III) measurement for $\theta<\theta_c$
($>\theta_c$), exhibiting a II--III ``transition'' and hence a cusp maximum at
$\theta=\theta_c$. The same holds for $I_1$ except that the transition between
the collinear and $Y$-type measurements is smoothed through an intermediate
region ($0.19\pi \alt \theta\alt 0.24\pi)$ where a {\it parity breaking}
measurement ($\gamma=0$, $0<\beta<\pi/4$ in (\ref{str1})) is preferred. These
features resemble then the $s=1/2$ case \cite{CRC.10,RCC.11}, where $D$
preferred a spin measurement along $x$ $\forall$ $\theta$ \cite{CRC.10} whereas
$I_2$ exhibited a sharp $z\rightarrow x$ transition, with  $I_1$ selecting a
parity breaking axis in a small intermediate interval \cite{RCC.11}. Hence, for
$s=1$, parity preserving type II and III measurements play the role of the $z$
and $x$ measurements respectively of the $s=1/2$ case.

Remarkably, the minimizing value of $\alpha$, obtained from Eq.\ (\ref{D11}),
{\it is the same} for $D$, $I_2$ and $I_1$ in all previous cases
$\forall\theta$ (i.e., for both type II and III measurements):
\begin{equation}
\tan\alpha=\tan^2\theta/2\,.\label{al}
\end{equation}
At this value the largest eigenvalue of $\rho'_{AB}$ is maximum and
$\rho'_{AB}$ attains certain majorizing properties. The evaluation of these
measures becomes then analytic. For instance, the quantum and geometric
discords read
\begin{eqnarray}
D&=&2h_{\frac{1}{2}}(p_\theta)-1-h_1[2q_\theta(1-q_\theta)]+h_1(q_\theta)\,,\\
I_2&=&\left\{^{\frac{1}{8}\sin^4\theta(3+\cos
2\theta)^2\,,\;\;\theta<\theta_c} _{\frac{1}{16}\cos^4\theta(11+4\cos 2\theta+\cos
4\theta)\,,\;\;\theta>\theta_c}\right.
 \,,\;\;\cos\theta_c={\textstyle\frac{1}{\sqrt[4]{3}}}\,,
 \end{eqnarray}
where $h_\nu(x)=-x\log_2 x-(\nu-x)\log_2(\nu-x)$, $p_\theta=\frac{1}{4}-
\frac{1}{16}(\frac{115}{8}-\cos2\theta+\frac{3}{2}\cos 4\theta+\cos
6\theta+\frac{1}{8}\cos 8\theta)^{1/2}$ and $q_\theta=\frac{1}{2}\sin^2\theta$.
Remarkably, $\theta_c\approx 0.23\pi$ in $I_2$ is determined by the overlap
condition $\langle -\theta|\theta\rangle^2=1/3$, as in the $s=1/2$ case
\cite{RCC.11}, with $I_2=2/9$ at $\theta=\theta_c$ (the same value as for
$s=1/2$). For $\theta\rightarrow 0$,  $D\approx \theta^2$ while $I_2\approx
2\theta^4$ (similar to the $s=1/2$ case \cite{CRC.10,RCC.11}), whereas for
$\theta\rightarrow\pi/2$, $D\approx[\frac{1}{2}-\frac{\log_2
e}{4}-\log_2(\frac{\pi}{2}-\theta)](\frac{\pi}{2}-\theta)^4$ while
$I_2\approx\frac{1}{2}(\frac{\pi}{2}-\theta)^4$. In this limit $D$ and $I_2$
are then proportional to the overlap $\langle
-\theta|\theta\rangle^{2}=\cos^{4s}\theta$. We also mention that for small
$\theta$, the difference between the approximate value of $D$ obtained with
spin measurements and the actual $D$ is very small ($O(-\theta^6\log_2\theta)$),
while in the case of $I_2$ and $I_1$, such difference is $O(\theta^4)$ (i.e.,
of leading order in $I_2$).

\section{Conclusions}
We have first provided a simple characterization of orthogonal projective
measurements in spin $1$ systems, which can be extended to arbitrary spin and
allows a rapid visualization of the (projective) measurements optimizing
discord-type measures of quantum correlations. Standard spin measurements are
not optimum in general for minimizing such measures for spin $s\geq 1$.
Instead, we have shown that for the relevant case of states with parity
symmetry, parity preserving measurements provide stationary solutions for all
these measures. We have identified such measurements for spin $1$, where they
are described by just two or three parameters (or one in the real case)
allowing to considerably simplify the variational problem associated with
discord. Results for the mixture (\ref{rhot}), which represents the state of
{\it any} spin pair in an XYZ chain in the immediate vicinity of the
factorizing field, confirm the optimality of such measurements in most cases.
They also confirm the distinct behavior of the minimizing measurement in the
quantum discord as compared to that in the geometric discord (or other measures
of type (\ref{If}) like the information deficit). The latter are more sensible
to changes in the nature of the state and hence more suitable for identifying
transitions between different regimes.

The authors acknowledge support of CIC (RR) and CONICET (JMM,NC) of Argentina.

\end{document}